\documentclass[aps,pre,twocolumn,showpacs,showkey,square,numbers,amssymb,amsmath,nobibnotes]{revtex4-1}
\usepackage{bm}
\usepackage{times,float}
\usepackage{graphicx}
\usepackage[usenames,dvipsnames,svgnames]{xcolor}
\usepackage{hyperref}
\hypersetup{colorlinks=true, linkcolor=NavyBlue, citecolor=PineGreen,urlcolor=cyan}

\usepackage{color}
\definecolor{dgreen}{rgb}{0,0.7,0}

\newcommand{\bracket}[1]{\left\langle #1\right\rangle}

\begin{document}
\title{Condensation of degrees emerging through a first-order phase transition in classical random graphs}
\author{Fernando L. Metz}
\address{Institute of Physics, Federal University of Rio Grande do Sul, 91501-970 Porto Alegre, Brazil}
\address{Physics Department, Federal University of Santa Maria, 97105-900 Santa Maria, Brazil}
\address{London Mathematical Laboratory, 8 Margravine Gardens, London W6 8RH, United Kingdom}
\author{Isaac P\'erez Castillo}
\address{Department of Quantum Physics and Photonics, Institute of Physics, UNAM, P.O. Box 20-364, 01000 Mexico City, Mexico}
\address{London Mathematical Laboratory,  8 Margravine Gardens, London W6 8RH, United Kingdom}

\begin{abstract}
 Due to their conceptual and mathematical simplicity, Erd\"os-R\'enyi or classical random graphs remain as a fundamental paradigm to model complex interacting systems in several areas. Although condensation phenomena have been widely considered in complex network theory, the condensation of degrees has hitherto eluded a careful study.  Here we show that the degree statistics of the classical random graph model undergoes a first-order phase transition between a Poisson-like distribution and a condensed phase, the latter characterized by a large fraction of nodes having degrees in a limited sector of their configuration space. The mechanism underlying the first-order transition is discussed in light of standard concepts in statistical physics. We  uncover the phase diagram characterizing the ensemble space of the model and we evaluate the rate function governing the probability to observe a condensed state, which shows that condensation of degrees is a rare statistical event akin to similar condensation phenomena recently observed in several other systems. Monte Carlo simulations confirm the exactness of our theoretical results.
\end{abstract}
\maketitle

\section{Introduction}
Condensation occurs when a macroscopic number of degrees of freedom occupy a small region of the configuration space. This is an ubiquitous phenomenon with manifestations in physics, biology, and economics \cite{Evans05}. Classical examples in physics are the familiar transition from a gas to a liquid state of matter, the formation of a Bose-Einstein condensate in large systems of non-interacting bosons \cite{Pathria}, or the emergence of a staggered magnetization in mean-field spin systems \cite{Berlin52,Cugliandolo95}. A crucial ingredient to observe condensation is the presence of global constraints, since they introduce correlations among the microscopic constituents of the system, and a condensed state can be formed even in the absence of interactions. In this context, Bose-Einstein condensation, where the total number of particles is conserved, constitutes the prototypical example.

In the physical examples mentioned above, condensation represents the typical or average behaviour of a large system in statistical equilibrium. More recently, it has been realized that condensation may take place in a broader variety of random systems when the ensemble space is probed away from the typical fluctuations around the average. The term {\it condensation of fluctuations} has been coined to describe such condensed states that are triggered by large deviations of an extensive random variable, but whose typical behaviour does not necessarily show any sign of condensation. Examples of random systems, where condensation emerges as a rare event, include the Gaussian model \cite{Zanetti14}, the Urn model \cite{Corberi15}, models of mass transport \cite{Evans06,Szavits14}, to name just a few. Condensation of fluctuations usually brings about  a rich phenomenology including phase transitions, giant responses to small perturbations, and singularities in the full probability distribution \cite{Corberi19}.

Here we study condensation of degree fluctuations in classical random graphs, introduced a long time ago by Solomonoff and Rapopport \cite{Solomon51}, and popularized a decade later by the seminal works of Erd\"os and R\'enyi \cite{Erdos59,Erdos60}. Due to their mathematical and conceptual simplicity, Erd\"os-R\'enyi (ER) random graphs have an enormous number of applications in areas such as complex networks, optimization problems, spin glasses, and information theory \cite{Mezard09,Newman10}. An instance drawn from the ER random graph ensemble consists of a simple undirected graph of $N$ nodes, where each pair of nodes is connected by an edge with probability $p \sim O(1/N)$ \cite{Newman10}. The degree $k_i$ is an integer random variable that counts the number of edges attached to node $i$, with $i=1,\dots,N$. Since the edges are drawn independently, the degree distribution along the graph follows a Poisson law in the large $N$ limit. Here we are precisely interested in condensation phenomena triggered by rare configurations of the random variables $\{ k_i \}_{i=1,\dots,N}$.

The study of condensation in the topological structure of complex networks or random graphs has attracted a lot of interest during the last two decades \cite{Doro08}. In this context, condensation refers to an aggregation phenomenon where a macroscopic number of elementary structures or motifs (edges, triangles, etc) become mutually interconnected to form a compact subgraph. Condensation of edges occurs when a finite fraction of links becomes attached to an infinitely small fraction of nodes \cite{Krap00,Bianconi01,Doro03,Doro05,Kartik14}, producing one or more hubs - densely connected nodes - in the graph structure. Condensation  has also been  observed in exponential random graphs \cite{Strauss86,Burda04a,Burda04b,Park04a,Park04b,Palla04,Park05,Annibale15}, which are sampled from a statistical weight with a Boltzmann form. The Hamiltonian of the exponential model is built in order to incorporate certain graph features, such as the total number of triangles \cite{Strauss86,Burda04a,Burda04b,Park05} or  two-stars \cite{Park04a,Park04b,Annibale15}, i.e., paths of length two.  Exponential random graphs also display a condensed state characterized by a large number of elementary structures (triangles or two-stars) clumped together into a highly interconnected cluster. These different manifestations of the condensed phase are typically characterized by large subgraphs with densely connected nodes, which translates in a subset of degrees scaling with $N$.

Here we take a more elementary path and study the condensation of the degree sequence $\{ k_i \}_{i=1,\dots,N}$ in a limited domain of its available configuration space.
Considering random graph instances drawn from the classical ER ensemble, we ask what is the fraction  of nodes having degrees inside an arbitrary interval $[a,b]$, with $b>a>0$. We provide an exact  solution to this problem by computing the full probability distribution of the random variable $F_N[a,b]$ that counts the fraction of degrees lying in $[a,b]$. Being more precise, by calculating the cumulant generating function of $F_N[a,b]$, we show that the ensemble of ER random graphs undergoes a first-order phase transition between a Poisson-like phase, where the degree distribution is closer to its typical behaviour, and a condensed phase, where the degree distribution exhibits a prominent peak. We elucidate the physical and mathematical mechanisms underlying the transition by using standard ideas from statistical mechanics. We derive the phase diagram in the ensemble parameter space and we show it exhibits two critical lines, each one surrounded by a metastable region and terminating at a critical point.
The critical lines define the set of points in the parameter space at which the degree distribution changes abruptly.
We also compute the rate function characterizing the large deviation probability of $F_N[a,b]$, whose striking property is the non-analytic behaviour. The calculation of the rate function shows that condensation of degrees is a rare statistical event in line with the condensation of fluctuations exhibited by other random systems \cite{Corberi19}. The theoretical results for the rate function exhibit an excellent agreement with Monte Carlo simulations.

In the next section we introduce the classical ensemble of random graphs and define the counting random variable $F_N[a,b]$. Section \ref{calc} explains the calculation of the cumulant generating function of  $F_N[a,b]$ using standard techniques of statistical mechanics. In section \ref{results} we present the results for the first-order condensation transition, the phase diagram, and the rate function. We summarize our results and conclusions in section \ref{conclusion}.

\section{The classical random graph model} 
\label{model}
The binary elements $c_{ij} \in \{ 0,1 \}$ of the $N \times N$ adjacency matrix defining the ensemble of Erd\" os-R\'enyi (ER) random graphs control whether there is an edge between pairs of nodes \cite{Newman10}: if $c_{ij}= 1$, nodes $i$ and $j$ are connected, while $c_{ij}= 0$ means there is no link between $i$ and $j$. Each instance of the ER ensemble is a simple and undirected graph  ($c_{ij} = c_{ji}$) without self-edges ($c_{ii} = 0$). The ensemble of ER random graphs can be defined by the following probability distribution for the adjacency matrix
\begin{equation}
P_{\rm ER}(\{ c_{ij} \}) = \prod_{i < j} \left[ \frac{c}{N} \delta_{c_{ij},1} + \left( 1 - \frac{c}{N}  \right) \delta_{c_{ij},0} \right],
 \label{jpqe}
\end{equation}  
where the product $\prod_{i < j}$ runs over all distinct pairs of nodes. The degree of node $i$, defined as
\begin{equation}
k_i^{(N)} = \sum_{j=1(\neq i)}^{N} c_{ij} ,
\label{degreee}
\end{equation}
gives the number of edges connected to $i$ in a single graph realization. The random variable $k_i^{(N)}$ fluctuates from node to node, and the average degree reads
\begin{equation}
c = \lim_{N \rightarrow \infty} \frac{1}{N} \sum_{i=1} \bracket{k_i^{(N)} },
\end{equation}
where $\langle \dots  \rangle$ represents the ensemble average over $\{ c_{ij} \}$ with the distribution in Eq. (\ref{jpqe}). Here we consider the sparse regime, where $c$ is finite and independent of $N$. Note that, in the above definition of the ER ensemble, $c$ is a control parameter that plays the role of a soft constraint on the degrees, changing dramatically the topological structure  of the random graphs generated from Eq. \eqref{jpqe}. Indeed, the ER  model undergoes a second-order percolation transition: for $c < 1$ the graph is composed of many finite clusters, each one containing a total number of $O(1)$ nodes, while for $c \geq 1$ a giant cluster with $O(N)$ nodes emerges continuously as a function of $c$ \cite{Newman10}.

Here we explore the condensation of degrees through the random variable
\begin{equation}
  F_N[a,b] = \frac{1}{N} \sum_{i=1}^{N} \left[ \Theta(b-k_i^{(N)}) -\Theta(a-k_i^{(N)})   \right],
  \label{kppqq}
\end{equation}  
where $\Theta(x)$ is the Heaviside step function. Clearly, $F_N[a,b]$  counts the fraction of nodes having degrees within the interval $[a,b]$. In the limit $N \rightarrow \infty$, its typical value becomes
\begin{equation}
  f_{\rm typ} \equiv \lim_{N \rightarrow \infty} \bracket{F_N[a,b]} = \sum_{k=0}^{\infty} p_c(k) I_{[a,b]}(k),
  \label{bbn}
\end{equation}
with $I_{[a,b]}(k) \equiv \Theta(b-k) -\Theta(a-k)$ an indicator function. The quantity $p_c(k)$ is the well-known $N \rightarrow \infty$ limit of the degree distribution of ER random graphs, given by a Poisson law with average $c$ \cite{Newman10}
\begin{equation}
  p_c(k) = \lim_{N \rightarrow \infty}  \frac{1}{N} \sum_{i=1}^{N} \bracket{\delta_{k,k_i^{(N)}} }= \frac{e^{-c} c^k}{k!},
  \label{poisson}
\end{equation}  
where the symbol $\delta$ denotes the Kronecker delta function. It is straightforward to check that $f_{ \rm typ}$ vanishes for $c \rightarrow \infty$ and $c \rightarrow 0$, whereas it has a maximum at some value of $c \in [a,b]$. Thus, when ER random graphs with $c \gg a,b$ are sampled from Eq. (\ref{jpqe}), the fraction $f_{\rm typ }$ is typically very small. However, in this particular situation, it is natural to ask what is the probability that Eq. (\ref{jpqe}) generates samples with a large subset of nodes with degrees in $[a,b]$, in spite of $c$ lying way outside $[a,b]$. Here we tackle this problem by computing exactly the full probability distribution of $F_N[a,b]$, which allows us to probe atypical, large fluctuations in the degree statistics of ER random graphs. We show that the ensemble space of ER random graphs have a surprisingly rich structure, displaying a first-order transition to a condensed configuration of  $\{ k_i \}_{i=1,\dots,N}$ caused by rare fluctuations of $F_N[a,b]$ around  its typical value.

\section{Calculation of the cumulant generating function} 
\label{calc}
The full statistics of $F_N[a,b]$ for large $N$ is captured by the cumulant generating function (CGF)
\begin{equation}
  \mathcal{G}_{[a,b]}(y) = \lim_{ N \rightarrow \infty} \frac{1}{N} \ln \mathcal{Z}_{[a,b]}^{(N)}(y),
  \label{kppc}
\end{equation}
where
\begin{equation}
  \mathcal{Z}_{[a,b]}^{(N)}(y) \equiv  \bracket{ e^{y N F_N[a,b] }  }.
  \label{hjqp}
\end{equation}  
All cumulants of $F_N[a,b]$ are obtained by taking derivatives of Eq. (\ref{kppc}) with respect to $y$. The leading contribution to the  probability $\mathcal{P}_{[a,b]}^{(N)}(f)$ of observing a fraction $0 \leq f \leq 1$ of nodes with degrees in $[a,b]$ decays, for large $N$, according to the large deviation principle \cite{touchette09}
\begin{equation}
  \mathcal{P}_{[a,b]}^{(N)}(f) \simeq \exp{\left[- N \Psi_{[a,b]} (f)   \right]} ,
  \label{ppo}
\end{equation}
where the rate function $\Psi_{[a,b]} (f)$ is the Legendre-Fenchel transform of $\mathcal{G}_{[a,b]}(y)$
\begin{equation}
  \Psi_{[a,b]} (f) = {\rm sup}_{y \in \mathbb{R}} \left[y f - \mathcal{G}_{[a,b]}(y)   \right].
  \label{rate}
\end{equation}  
There is a natural analogy between the canonical ensemble of statistical mechanics and the framework introduced  above. The control parameter $y$ plays the role of inverse temperature, $y  \mathcal{G}_{[a,b]}(y)$ is the free-energy per degree of freedom, and $\mathcal{Z}_{[a,b]}^{(N)}(y)$ is the partition function. As we will show below, the behaviour of $\mathcal{G}_{[a,b]}(y)$ and its derivative allows to clearly identify a first-order phase transition. As can be noted from Eq. (\ref{hjqp}), $y$ is responsible for  biasing the configurations of the ER ensemble: positive (negative) values of $y$ favour configurations corresponding
to large (small) values of $F_N[a,b]$.
Thus, for a fixed value of $y$, we have a measure from which to derive the relevant statistical properties of the biased ensemble describing atypical graph realizations. In particular, the
degree distribution of the constrained ensemble is simply given by:
\begin{equation}
p_y(k)=\lim_{N\to\infty}\frac{\bracket{\frac{1}{N}\sum_{i=1}^N\delta_{k,k_i}e^{y N F_N[a,b]}}}{\bracket{e^{y N F_N[a,b]}}}\,.
\label{eq:edd}
\end{equation}

Let us proceed to the calculation of the CGF. The partition function can be rewritten as
\begin{equation}
   \mathcal{Z}_{[a,b]}^{(N)}(y) = \sum_{k_1,\dots,k_N=0}^{N-1} e^{y \sum_{i=1}^N I_{[a,b]}(k_i)  }
    \mathcal{P}_N(k_1,\dots,k_N) ,
\end{equation}  
where
\begin{equation}
\mathcal{P}_N(k_1,\dots,k_N) \equiv  \left \langle \prod_{i=1}^{N}  \delta_{k_i,  \sum_{j=1 }^N c_{ij}} \right \rangle
\end{equation}  
is the joint distribution of degrees. Note that the degrees at different nodes are correlated random variables. By introducing the integral representation of the Kronecker delta function
\begin{equation}
\delta_{k_i,  \sum_{j=1}^N c_{ij}} = \int_{0}^{2 \pi} \frac{d u_i}{2 \pi} \exp{\left( i u_i k_i - i u_i \sum_{j=1 (\neq i)}^N c_{ij}   \right) } , 
\end{equation}  
we obtain
\begin{align}
 & \mathcal{Z}_{[a,b]}^{(N)}(y) = \sum_{k_1,\dots,k_N=0}^{N-1}
    \int \left( \prod_{i=1}^{N} \frac{d u_i}{2 \pi}   \right)
    \exp{\left( i \sum_{i=1}^{N} u_i k_i \right) }
    \nonumber \\
    &\times
 \exp{\left[  y \sum_{i=1}^N I_{[a,b]}(k_i) \right] }
 \left \langle   \exp{\left[ - i \sum_{i < j} c_{ij} (u_i + u_j)   \right]}  \right \rangle .
  \label{kpq}
\end{align}  
The ensemble average in Eq. (\ref{kpq}) is readily computed, leading to the following expression for large $N$
\begin{align}
 & \mathcal{Z}_{[a,b]}^{(N)}(y) = \sum_{k_1,\dots,k_N=0}^{N-1}
    \int \left( \prod_{i=1}^{N} \frac{d u_i}{2 \pi}   \right)
    \exp{\left( i \sum_{i=1}^{N} u_i k_i \right) }
    \nonumber \\
    &\times
 \exp{\left[  y \sum_{i=1}^N I_{[a,b]}(k_i)  - \frac{c N}{2}  + \frac{c}{2 N} \left( \sum_{i=1}^N e^{-i u_i} \right)^2  \right]   }\,,
  \label{kpq11}
\end{align}  
where we have retained only the leading terms of $O(N)$ in the exponent. The above equation couples the variables $u_i$ on different
nodes, which prevents the calculation of the integrals over $u_i$. However,  by employing the Hubbard-Stratonovich transformation
\begin{equation}
\int_{-\infty}^{\infty} d \mu \exp{\left( - \frac{1}{2} N c \mu^2 + c J \mu  \right)} = \sqrt{\frac{ 2 \pi }{ c N  }} \exp{\left( \frac{c}{2N} J^2  \right)}
\label{id}
\end{equation}  
in Eq. (\ref{kpq11}), with $J= \sum_{i=1}^N e^{-i u_i}$, we are able to integrate over $u_1,\dots,u_N$ and recast the CGF in integral form

\begin{equation}
  \mathcal{G}_{[a,b]}(y) = \lim_{N \rightarrow \infty} \frac{1}{N} \ln \left[  \int_{-\infty}^{\infty} d \mu \, e^{ N \mathcal{F}_{[a,b]}(\mu|y) }
    \right] ,
  \label{kpqq}
\end{equation}  
where
\begin{equation}
  \mathcal{F}_{[a,b]}(\mu|y) =   \frac{c }{2}  - \frac{1}{2}  c \mu^2 + \ln{\left( \sum_{k=0}^{\infty} p_c(k) e^{y I_{[a,b]}(k)} \mu^k    \right)    }.
  \label{jjbb}
\end{equation}

In the limit $N \rightarrow \infty$, the integral in Eq. (\ref{kpqq}) is dominated by the {\it global maximum} of $\mathcal{F}_{[a,b]}(\mu|y)$ with respect to the order-parameter $\mu$, and the integral can be solved through the Laplace method. Defining $\mu_{g}$ as the global maximum of $\mathcal{F}_{[a,b]}(\mu|y)$, we obtain
\begin{equation}
  \mathcal{G}_{[a,b]}(y) = \mathcal{F}_{[a,b]}(\mu_g|y),
  \label{jjj}
\end{equation}  
where $\mu_g$ is determined from the solution of the fixed-point equation 
\begin{equation}
  \mu = \frac{\sum_{k=0}^{\infty} p_c(k)   e^{y I_{[a,b]}(k+1)} \mu^k    }{\sum_{k=0}^{\infty} p_c(k)   e^{y I_{[a,b]}(k)} \mu^k    }
  \label{jqn}
\end{equation}  
that corresponds to the global maximum of the function $\mathcal{F}_{[a,b]}(\mu|y)$ given by Eq. \eqref{jjbb}. The fixed-point equation  \eqref{jqn} is derived by requiring that $\mathcal{F}_{[a,b]}(\mu|y)$ is stationary with respect to $\mu$, i.e., $\frac{d \mathcal{F}_{[a,b]}(\mu|y) }{d \mu}  = 0$.
In principle,  $\mathcal{F}_{[a,b]}(\mu|y)$ may have more than a single maximum depending on the control parameters $(c,y)$, which translates in more than a single fixed-point solution to Eq. (\ref{jqn}). The rate function follows from Eq. (\ref{rate})
\begin{equation}
  \Psi_{[a,b]} (f) = y f - \mathcal{F}_{[a,b]}(\mu_g|y),
  \label{rate1}
\end{equation}  
where the fraction $0 \leq f \leq 1$ is obtained from the first derivative of the CGF $f = \frac{\partial \mathcal{F}_{[a,b]}(\mu_g|y) }{ \partial y}$, namely
\begin{equation}
  f   = \frac{  \sum_{k=0}^{\infty} p_c(k)I_{[a,b]}(k) e^{y I_{[a,b]}(k)} \mu_g^k     }{  \sum_{k=0}^{\infty} p_c(k) e^{y I_{[a,b]}(k)} \mu_g^k    }.
  \label{hja}
\end{equation}  
In a similar manner, one can show that the degree distribution of the constrained ensemble, defined in Eq. (\ref{eq:edd}), takes the following form:
\begin{equation}
  p_y(k) = \frac{p_c(k)  e^{y I_{[a,b]}(k)} \mu_g^k     }{  \sum_{k=0}^{\infty} p_c(k) e^{y I_{[a,b]}(k)} \mu_g^k    }.
 \label{eq:edd1}
\end{equation}
For $y=0$, we obtain $\mu=1$, which implies that $f$ and $p_y(k)$ reduce to their standard expressions arising from typical fluctuations of ER random graphs (see eqs. (\ref{bbn}) and (\ref{poisson})).
Equations (\ref{jjj}-\ref{eq:edd1}) constitute the main analytical results of this work, as they determine completely the statistical properties of the random variable $F_N[a,b]$  for ER random graphs.

\section{First-order transition and condensation of degrees } 
\label{results}

\begin{figure}
\begin{center}
  \includegraphics[height=6cm,width=8cm]{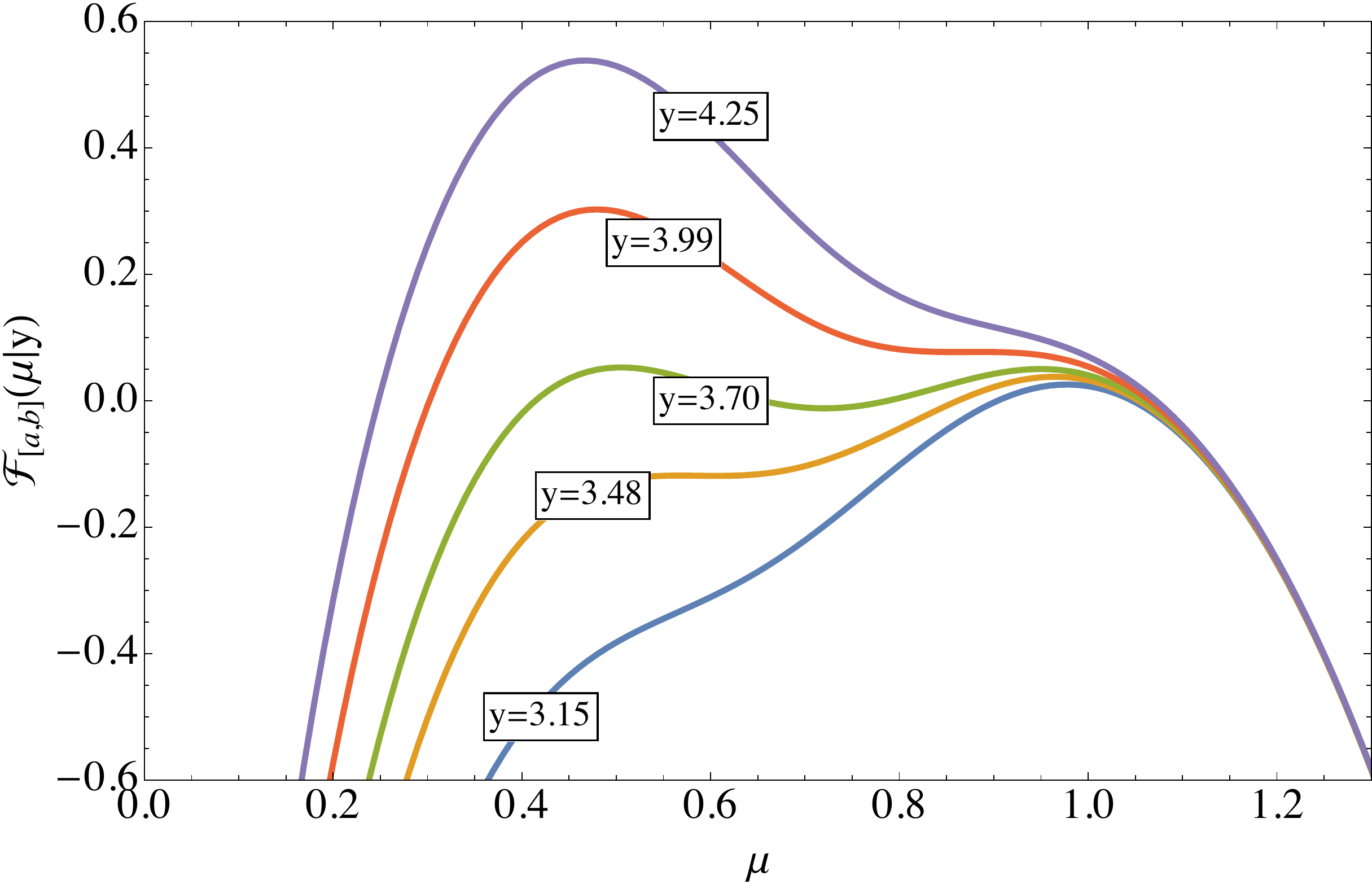}
  \caption{The quantity $\mathcal{F}_{[a,b]}(\mu|y)$ as a function of the order-parameter $\mu$ for average degree $c=13$, $a=1$, $b=3$, and increasing values of $y$. The global maximum of $\mathcal{F}_{[a,b]}(\mu|y)$ with respect to $\mu$ provides the cumulant generating function (see Eqs. (\ref{kpqq}-\ref{jjj})). The behaviour of $\mathcal{F}_{[a,b]}(\mu|y)$ characterizes the emergence of a first-order transition: for $y$ approximately  in the range $(3.48,3.99)$, the function $\mathcal{F}_{[a,b]}(\mu|y)$ exhibits two maxima, which have the same height only at the critical value $y_{\rm FT } \simeq 3.70$. The function  $\mathcal{F}_{[a,b]}(\mu|y)$ has a single maximum for $y \lesssim 3.48$ and $y \gtrsim 3.99$.
}
\label{CGFvaryy}
\end{center}
\end{figure}

In order to understand the mechanism underlying the first-order transition, we start investigating the behaviour of $\mathcal{F}_{[a,b]}(\mu|y)$ as a function of the order-parameter $\mu$. The function $\mathcal{F}_{[a,b]}(\mu|y)$ plays the analogous role as the functional free-energy in the canonical ensemble of mean-field models. Here the global maximum  $\mathcal{F}_{[a,b]}(\mu_g|y)$ yields the CGF, which is
analogous to the (equilibrium) free-energy in statistical mechanics.
Figure \ref{CGFvaryy} depicts the typical functional form of $\mathcal{F}_{[a,b]}(\mu|y)$ for high $c$ and increasing values of $y$.
As we can notice, for small $y$, closer to the typical case $y=0$,  $\mathcal{F}_{[a,b]}(\mu|y)$ has a single maximum. Increasing the value of $y$ leads to the emergence of a second maximum, which means that Eq. (\ref{jqn}) admits three fixed-point solutions: two maxima and one minimum. The portion of the phase diagram where $\mathcal{F}_{[a,b]}(\mu|y)$ has two maxima defines a metastable region, within which the global maximum yields the leading contribution to the integral in Eq. (\ref{kpqq}), while the other maximum corresponds to a metastable state, providing a sub-leading term to the CGF. Finally, the heights of $\mathcal{F}_{[a,b]}(\mu|y)$ corresponding to the two different maxima become even only at $y_{FT}$, and both fixed-point solutions for $\mu$ contribute equally to the saddle-point integral in Eq. (\ref{kpqq}).   The set of critical values $(c_{\rm FT},y_{\rm FT})$ in the parameter space, where both solutions contribute equally to the CGF, defines a first-order transition line.

\begin{figure}[h!]
\begin{center}
  \includegraphics[height=6cm,width=8cm]{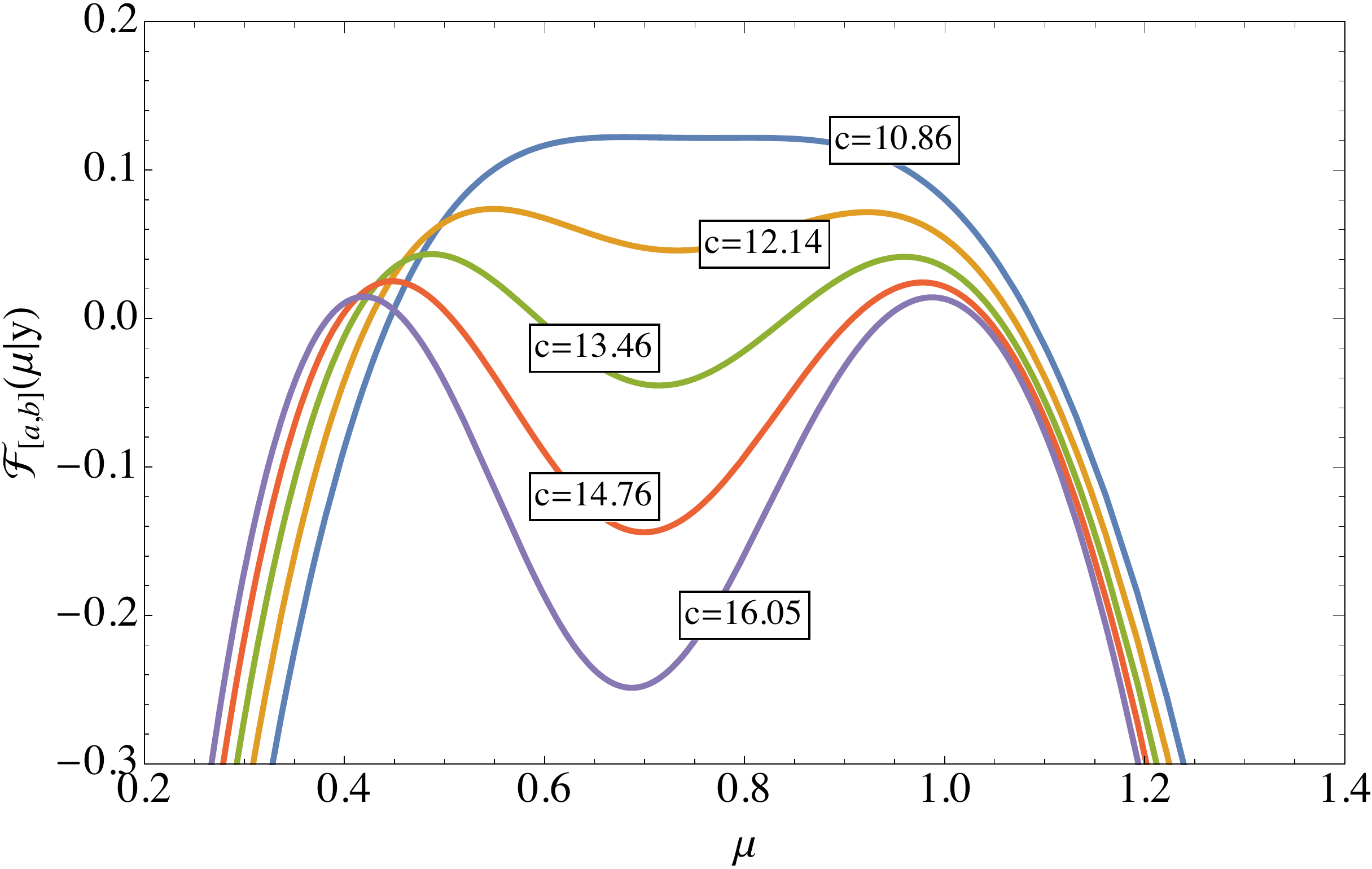}
  \caption{The quantity $\mathcal{F}_{[a,b]}(\mu|y)$ as a function of the order-parameter $\mu$ for $a=1$, $b=3$, and different values of the average degree $c$ along the upper critical  line (see figure \ref{dens}). The two maxima have the same height and thus contribute equally to the cumulant generating function, which features the coexistence of phases. The two maxima merge into a single maximum right at the critical point.}
\label{coexistence}
\end{center}
\end{figure}

Figure \ref{coexistence} also illustrates $\mathcal{F}_{[a,b]}(\mu|y)$ as a function of $\mu$ for some values of $(c_{\rm FT},y_{\rm FT})$.
As we move along the critical line in the parameter space, the two maxima of $\mathcal{F}_{[a,b]}(\mu|y)$ gradually  approach each other, until they  finally merge into a single maximum at a critical point, marking the end of the critical line. This physical picture is analogous to the Van der Waals liquid-gas phase transition or the ferromagnetic mean-field transition in the presence of an external field, with the proviso that our current exact analysis is not a mean field theory for a more complex underlying model. As a consequence, the metastable region cannot be promoted to a coexistence region.

\begin{figure}[h!]
\begin{center}
  \includegraphics[height=6cm,width=8cm]{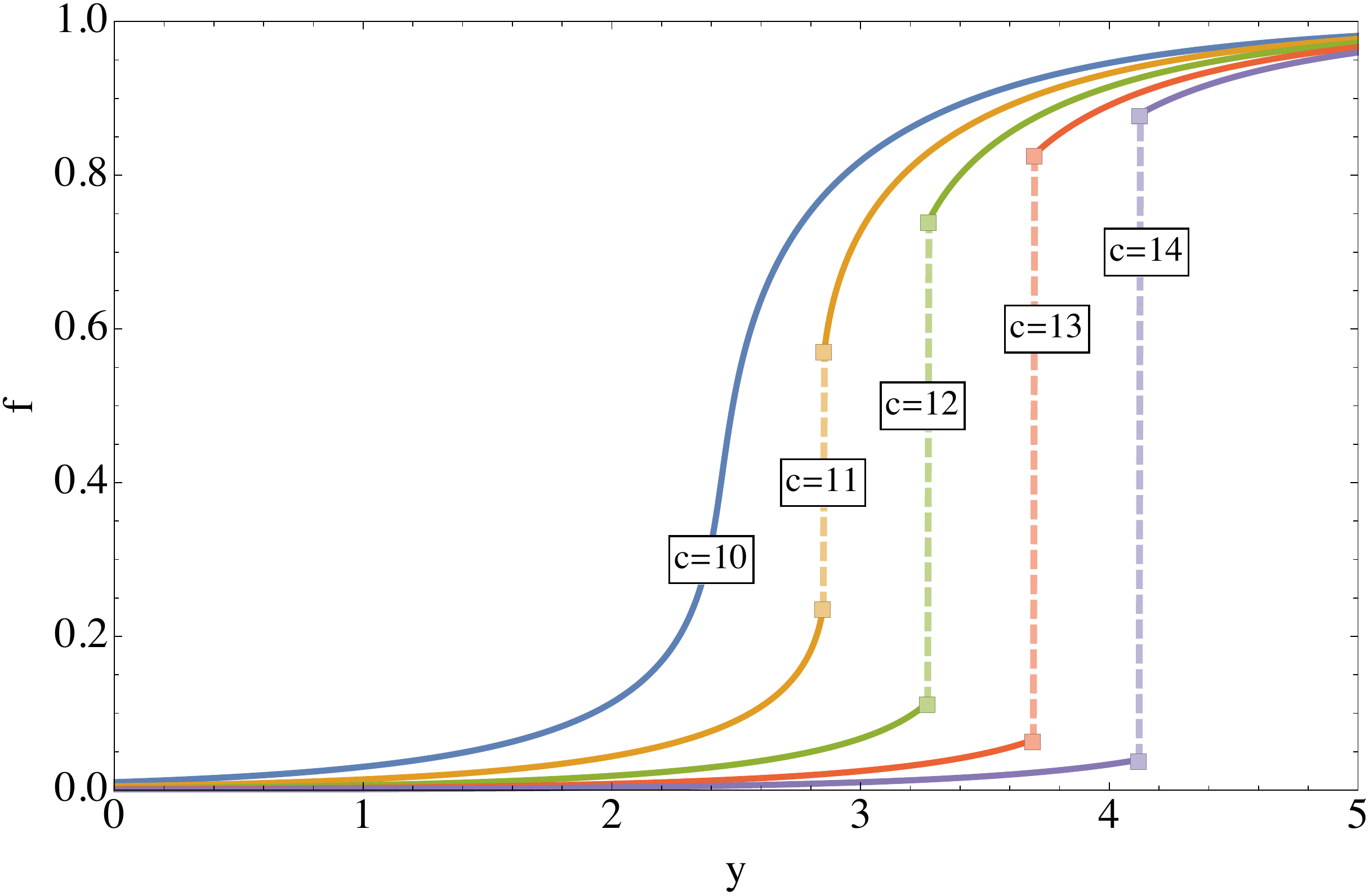}
\caption{Fraction $f$ of degrees within the interval $[1,3]$ as a function of $y$ for different values of the average degree $c$. For large values of $c$, $f$ has a discontinuous behaviour at a critical value $y_{\rm FT}$, marking the first-order phase transition.}
\label{fraction}
\end{center}
\end{figure}

In light of the aforementioned discussion, the CGF $\mathcal{G}_{[a,b]}(y)$, obtained from the global maximum $\mathcal{F}_{[a,b]}(\mu_g|y)$, is necessarily a continuous function of $(c,y)$. However, the derivative of $\mathcal{G}_{[a,b]}(y)$ with respect to $y$, which renders the fraction $f$, exhibits a jump when crossing the transition line. Figure \ref{fraction} shows the discontinuous behaviour of $f$ as a function of $y$ which emerges at sufficiently large values of $c$ in comparison to the interval $[a,b]$. Such discontinuous behaviour of the first derivative of $\mathcal{G}_{[a,b]}(y)$ is the hallmark of a {\it first-order phase transition}. As illustrated in figure \ref{fraction}, the discontinuity becomes more prominent for increasing $c \gg 1$, whereas below a certain value of $c$ the fraction $f$ increases smoothly with $y$.

\begin{figure}[h!]
\begin{center}
  \includegraphics[height=6cm,width=8cm]{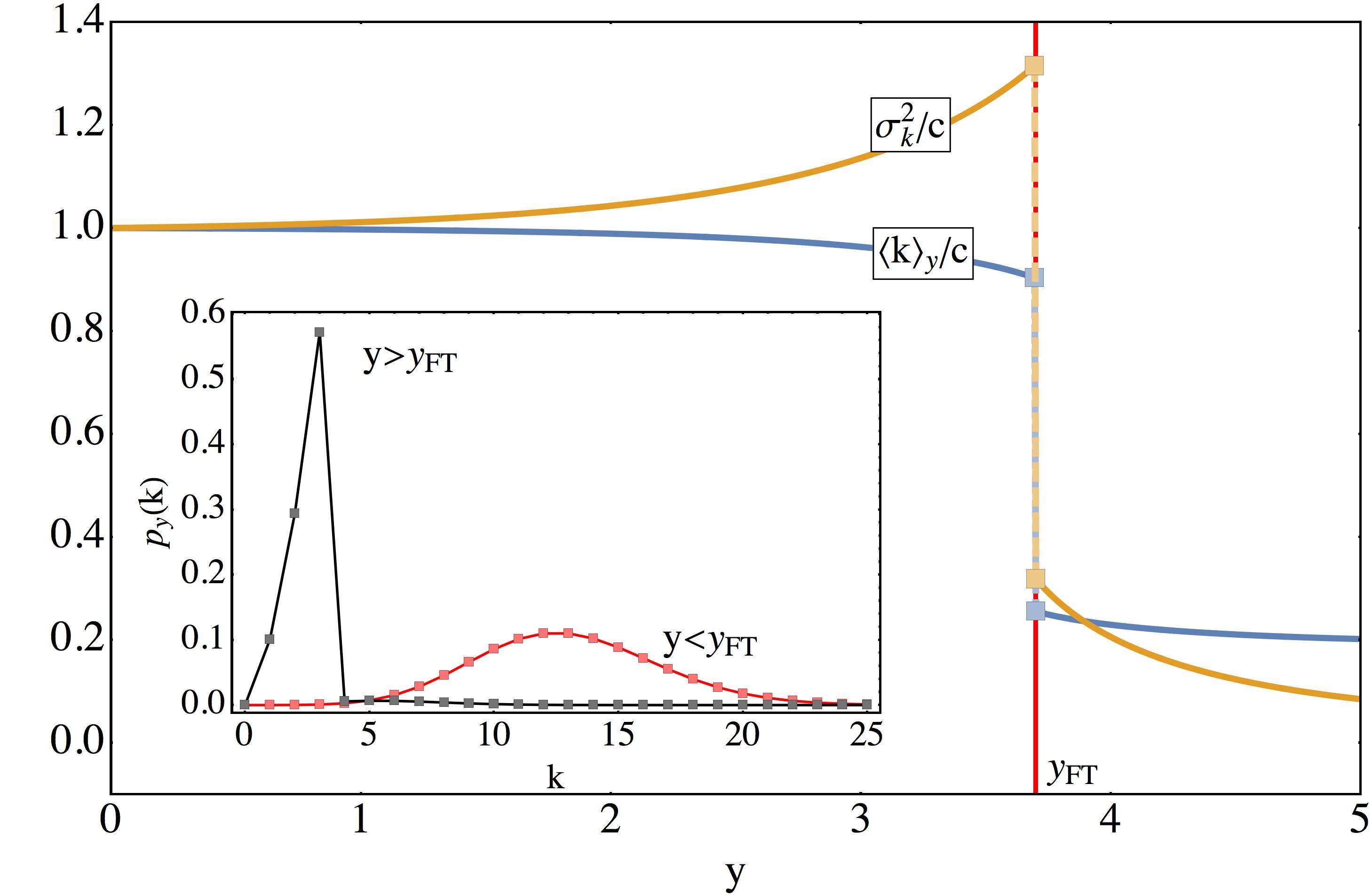}
\caption{ Average $\langle k \rangle_y$ and variance $\sigma_{k}^{2}$ of the degree distribution $p_y(k)$ (see Eq. \eqref{eq:edd}) characterizing rare graph samples generated from  Eq. (\ref{jpqe}), with $c=13$, and conditioned to have a certain fraction $f$ of degrees inside the interval $[1,3]$.  The concurrent behaviour of $f$ as a function of $y$ is presented in Fig. \ref{fraction}. The inset shows the typical profile of the degree distribution in each phase. The quantities $\langle k \rangle_y$ and $\sigma_{k}^{2}$ have a discontinuous behaviour that features the abrupt change of the degree statistics along the first-order phase transition.}
\label{degreedistr}
\end{center}
\end{figure}

Let us now characterize the different phases by studying the behaviour of the degree distribution $p_y(k)$ along the transition line.  Figure \ref{degreedistr} 
shows the average $\langle k \rangle_y$  and the variance $\sigma_{k}^{2} = \langle k^2 \rangle_y - \langle k \rangle_{y}^{2}$ of $p_y(k)$ for $c=13$ and increasing
values of $y$. Clearly, the degree statistics changes abruptly at the critical point $y_{\rm FT}$, with both $\langle k \rangle_y$ and $\sigma_{k}^{2}$ exhibiting a discontinuous behaviour. For $y < y_{\rm FT}$, the random graph instances have a wider range of degrees and the distribution $p_y(k)$ is closer to a Poisson law with average $c$. For $y > y_{\rm FT}$, the random graph instances become more homogeneous, since the majority of nodes have degrees within $[a,b]$. Accordingly, the variance $\sigma_{k}^{2}$ becomes smaller than $\langle k \rangle_y$ and the distribution $p_y(k)$ has a peak at a certain $k \in [a,b]$, closer to the degree distribution of a random regular graph.
In fact, if we choose an interval $[a,b]$ such that it contains a single degree $K$,  Eq. (\ref{eq:edd}) converges to $p_y(k) = \delta_{k,K}$ for $y \rightarrow \infty$.
The degree distribution $p_y(k)$ in each phase is shown as an inset in figure \ref{degreedistr}.
In summary, ER random graphs undergo a topological first-order transition between an heterogeneous phase, identified by a broader spectrum of degrees, to an homogeneous phase, where the degrees condensate in the interval $[a,b]$.

\begin{figure}[h!]
\begin{center}
  \includegraphics[height=8cm,width=8cm]{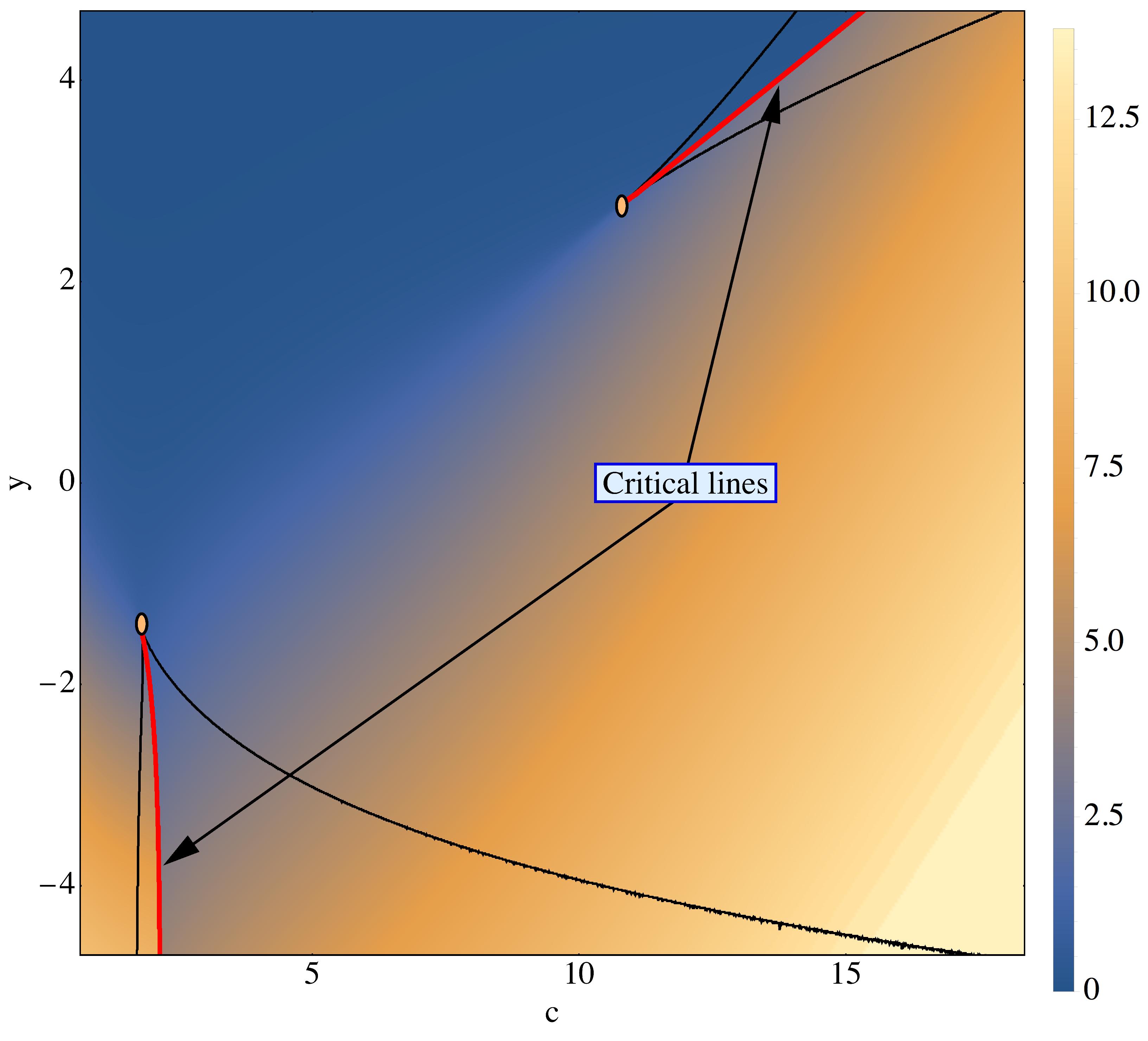}
  \caption{Phase diagram showing  the fraction $f$ of degrees lying in $[1,3]$ for each combination of parameters $(c,y)$. The values of $-\log(f)$ are displayed in a density plot according to the colour scale. The red curves denote first-order transition lines. The black curves delimit the regions of the phase diagram where the saddle-point Eq. (\ref{jqn}) has three fixed-point solutions. The solid yellow circles represent critical points at which the two maxima of $\mathcal{F}_{[a,b]}(\mu|y)$ merge into a single one (see figure \ref{coexistence}).
}
\label{dens}
\end{center}
\end{figure}

The above results are summarized in the phase diagram of figure \ref{dens}, where a density plot for the fraction $f$ on the parameter space $(c,y)$ is presented in a logarithmic colour scale.
As we can see, there exists two first-order critical lines, indicated by solid red lines, for small and large values of the average degree $c$. Each critical line terminates at a critical point (solid red circles).
The black curves delimit the metastable regions around each first-order critical line, within which Eq. (\ref{jqn}) has three fixed-point solutions. 
We remind the reader that Eq. (\ref{jpqe}) produces random graph samples with average degree $c$, while positive (negative) values of $y$ favour random graph configurations with an average fraction $f$ of degrees in $[a,b]$ larger (smaller) than its typical value. As we can appreciate from figure \ref{dens}, by fixing $y > 0$ sufficiently large, the first-order transition appears for $c$ large in comparison to $[a,b]$, when we simultaneously require that samples have a large average degree and a large fraction $f$. By setting $y < 0$, with $|y|$  sufficiently large, the first-order transition appears for $c \in [a,b]$, which is incompatible with a small fraction $f$.  Thus, the appearance of two first-order phase transitions is due to the existence of two distinct situations where conflicting constraints are imposed on the generation of random graph samples.

\begin{figure}[ht!]
\begin{center}
\vspace{.5cm}
\includegraphics[height=6cm,width=8cm]{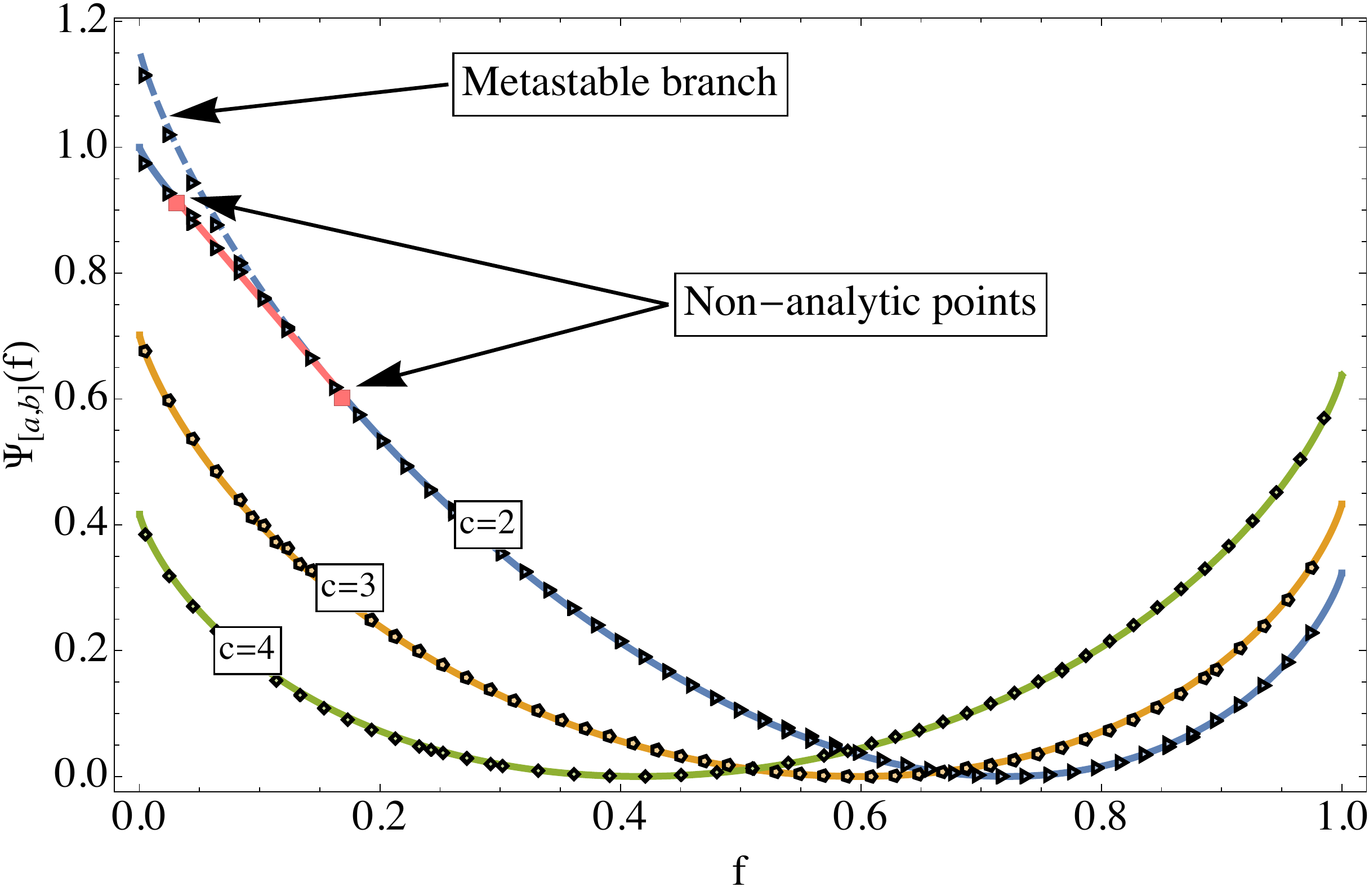}
\caption{Rate function $\Psi_{[a,b]} (f)$, characterizing the large deviation probability of Eq. (\ref{ppo}), as a function of the fraction $f$ of nodes with degrees within the interval $[1,3]$. We show the rate function for different values of the average connectivity $c$. The function $\Psi_{[a,b]} (f)$ has a minimum at the typical value $f=f_{\rm typ}$ (see Eq. (\ref{bbn})). For the case $c=2$ (solid blue line), we also show the presence of two non-analytic points (light red squares) at which the derivatives of the rate function are discontinuous. By the construction of the Legendre-Fenchel transform \cite{touchette}, these points are connected by a straight line (solid light red line). The dashed blue line is the rate function obtained from choosing the metastable solution for the order-parameter $\mu$ when calculating the CGF. The different symbols (triangles, pentagons and rhombuses) are simulation results obtained through a reweighting Monte Carlo method.}
\label{ratefunc}
\end{center}
\end{figure}

Finally, we compute the rate function $\Psi_{[a,b]} (f)$ controlling the large deviation probability of having values of $f$ away from its typical value $f_{\rm typ}$ (see Eq. (\ref{ppo})). According to Eq. \eqref{rate}, the rate function is the Legendre-Fenchel transform of the CGF with respect to $y$. Thus, the fact that for certain values of $c$ the CGF has a non-analytic behaviour in its
first derivative $\frac{d \mathcal{G}_{[a,b]}(y)}{dy}$ results in the appearance of a non-analytical behaviour on the rate function. This is confirmed by Fig. \ref{ratefunc}, where we show $\Psi_{[a,b]} (f)$ as a function of $f$ for different values of $c$. For $c=2$, the rate function presents two non-analytic points, connected by a straight line, at which the derivatives from the left and the right of each non-analytic point present a jump. Since the condensed phase is characterized  by a very small fraction $f$ in the case of $c=2$, figure  \ref{ratefunc} clearly shows that condensation of degrees is a rare statistical event. By expanding the rate function around its minimum $\Psi_{[a,b]} (f_{\rm typ}) = 0$,  we obtain 
\begin{equation}
\Psi_{[a,b]} (f) = \frac{1}{2} \frac{\left( f - f_{\rm typ}  \right)^2}{\sigma_f^2},
\end{equation}
with
\begin{equation}
\sigma_f^2 = c \left[ \sum_{k=0}^{\infty} p_c(k) I_{[a,b]}(k+1) - f_{\rm typ}  \right]^2  + f_{\rm typ} \left( 1- f_{\rm typ} \right).
\end{equation}  
This implies that the typical fluctuations of $f$ around $f_{\rm typ}$ are described by a Gaussian distribution with variance $\sigma_f^2$.

In order to confirm the exactness of our theoretical findings, we have performed a reweighting  Monte Carlo method to estimate the rate function \cite{Hartmann2018}. In our particular case, one must be careful when estimating the rate function for parameters $(c,y)$ within the metastable region, as the Monte Carlo simulation may be trapped in the metastable solution. However, this problem is easily surmounted by choosing the appropriate initial conditions. The simulation results and their comparison with our theoretical findings are also shown in Fig. \ref{ratefunc}, where we have also included the estimation of the rate function obtained from the metastable branch. The comparison between our theory and simulations shows a very good agreement.

\section{Conclusions} 
\label{conclusion}
In this work we have studied the large deviation properties of the degree sequence characterizing the Erd\"os-R\'enyi ensemble of random graphs. By studying the fluctuations of an elementary observable, namely the fraction $f$ of degrees lying in a given interval, we have shown that the ensemble space of ER random graphs exhibit rich critical phenomena, with the presence of two first-order critical lines marking a topological transition in the degree statistics. As the transition lines are crossed, we have shown that the degree distribution changes abruptly from a Poisson-like profile, characteristic of typical samples from the ER ensemble, to a peaked distribution. The latter degree distribution identifies  a phase where degrees are condensed or concentrated in a limited sector of their configuration space and, consequently, random graph samples corresponding to this phase have a rather homogeneous structure, similar to regular random graphs. Nevertheless, condensation of degrees in the ER ensemble is an extremely rare event, as confirmed by our computation of the rate function describing the large deviation probability for the fraction $f$. Such rate function exhibits two non-analytic points akin to the presence of the first-order transition. We point out that similar properties have been identified in the probability distributions describing the condensation of fluctuations in other disordered systems \cite{Zanetti14,Corberi19}. Our theoretical results for the rate function are fully confirmed by Monte Carlo simulations.

There are an interesting number of questions to be explored. Firstly, one wonders whether the metastable region might become a coexistence phase for a different system for which our solution would correspond to its mean-field approximation. In this regard studies on  Euclidean graphs \cite{Gilbert1961} seem to be the most natural candidate. Secondly, in the same manner that the percolation transition is inherited in the magnetic properties of the Ising model on random graphs, it is pertinent to ask what is the impact on the magnetic properties of the topological first order transition we have observed here. In a similar context, it would be also interesting to study whether large deviations in the degree sequence can trigger the decay of the metastable states found in coupled ER random graphs \cite{Maira2018}. Finally, in light of recent advances in the study of large deviations on diluted random matrices \cite{Perez2016,Perez2018,Perez2018b}, we should consider the spectral properties corresponding to the constrained graph ensemble studied here. Some of these questions are currently under consideration. As a last remark, we point out that the techniques introduced here can be also useful to study exponential random graphs \cite{Newman10,Park04a,Annibale15}, since the approach of section \ref{calc} can be readily generalized to the case where the indicator function is replaced by any other arbitrary function of a single degree.

\begin{acknowledgments}
FLM and IPC thank London Mathematical Laboratory for financial support. FLM also acknowledges a fellowship and financial support from
CNPq/Brazil (Edital Universal 406116/2016-4).
\end{acknowledgments}

\bibliography{biblio}

\end{document}